\pdfoutput=1
\documentclass[11pt,a4paper]{article}
\usepackage{jcappub}

\usepackage{bm}
\usepackage{amsfonts}
\usepackage{subfigure}

\renewcommand\({\left(}
\renewcommand\){\right)}
\renewcommand\[{\left[}
\renewcommand\]{\right]}

\newcommand{\GF}{G_{\rm F}}
\newcommand{\I}{{\rm i}}
\newcommand{\half}{{\textstyle\frac{1}{2}}}

\long\def\dump#1{}

\begin{document}


\title{Self-induced neutrino flavor conversion without flavor mixing}

\author[a]{S.~Chakraborty,}
\author[b]{R.~S.~Hansen,}
\author[a]{I.~Izaguirre,}
\author[a]{and G.~G.~Raffelt}

\affiliation[a]{Max-Planck-Institut f{\"u}r Physik
(Werner-Heisenberg-Institut),\\
F{\"o}hringer Ring 6, 80805 M{\"unchen}, Germany}

\affiliation[b]{Max-Planck-Institut f{\"u}r Kernphysik,\\
Saupfercheckweg 1, 69117 Heidelberg, Germany}

\emailAdd{sovan@mpp.mpg.de}
\emailAdd{rasmus@mpi-hd.mpg.de}
\emailAdd{izaguirr@mpp.mpg.de}
\emailAdd{raffelt@mpp.mpg.de}

\abstract{Neutrino-neutrino refraction in dense media can cause
  self-induced flavor conversion triggered by collective run-away
  modes of the interacting flavor oscillators. The growth rates were
  usually found to be of order a typical vacuum oscillation frequency
  $\Delta m^2/2E$. However, even in the simple case of a $\nu_e$ beam
  interacting with an opposite-moving $\bar\nu_e$ beam, and allowing
  for spatial inhomogeneities, the growth rate of the fastest-growing
  Fourier mode is of order $\mu=\sqrt{2} \GF n_{\nu}$, a typical
  $\nu$--$\nu$ interaction energy. This growth rate is much larger
  than the vacuum oscillation frequency and gives rise to flavor
  conversion on a much shorter time scale. This phenomenon of ``fast
  flavor conversion'' occurs even for vanishing $\Delta m^2/2E$ and
  thus does not depend on energy, but only on the angle
  distributions. Moreover, it does not require neutrinos to mix or to
  have masses, except perhaps for providing seed disturbances.  We
  also construct a simple homogeneous example consisting of
  intersecting beams and study a schematic supernova model proposed by
  Ray Sawyer, where $\nu_e$ and $\bar\nu_e$ emerge with different
  zenith-angle distributions, the key ingredient for fast flavor
  conversion.  What happens in realistic astrophysical scenarios
  remains to be understood.}

\maketitle

\section{Introduction}
\label{sec:introduction}

Most of the energy liberated in stellar core collapse or in neutron-star
mergers appears in the form of neutrinos and antineutrinos of all flavors,
but with fluxes and spectra that differ strongly between $\nu_e$, $\bar\nu_e$
and the other species, collectively referred to as $\nu_x$. The subsequent
flavor evolution of these neutrinos influences energy deposition beyond the
decoupling region, neutrino-driven nucleosynthesis, and detection
opportunities of the neutrino signal from the next nearby supernova or the
diffuse supernova neutrino flux from all past core-collapse events
\cite{Janka:2012wk, Scholberg:2012id, Mirizzi:2015eza}. However, a true
understanding of flavor evolution in dense environments has remained elusive
because of many complications engendered by the nonlinear nature of
collective flavor oscillations \cite{Mirizzi:2015eza, Chakraborty:2016yeg}. We study a
new item on this list which has eluded most workers in this field with the
notable exception of Ray Sawyer \cite{Sawyer:2005jk, Sawyer:2015dsa}, i.e.,
the surprising insight that collective flavor conversion need not depend on
neutrino mixing parameters.

Collective neutrino flavor oscillations manifest themselves in the form of
two generic phenomena. One is the effect of synchronisation: different modes
of the neutrino mean field oscillate in lockstep even though they have
different vacuum oscillation frequencies $\omega=\Delta m^2/2E$
\cite{Kostelecky:1994dt, Pantaleone:1998xi, Pastor:2001iu, Raffelt:2010za,
Akhmedov:2016gzx}. The other is the phenomenon of self-induced flavor
conversion, corresponding to collective run-away modes
\cite{Kostelecky:1993dm, Samuel:1995ri, Duan:2005cp, Hannestad:2006nj,
Duan:2007mv, Banerjee:2011fj, Raffelt:2011yb}. Surprisingly, the growth rate
in the linear regime and the overall evolution need not depend on $\Delta
m^2/2E$ and therefore the effect can occur even for unmixed neutrinos if
given an appropriate seed to grow from \cite{Sawyer:2005jk, Sawyer:2015dsa}.

Collective ``flavor conversion'' actually does not represent any change of
flavor in the overall ensemble, but a reshuffling among different modes. In
the simplest case, a gas of $\nu_e$ and $\bar\nu_e$ can convert to $\nu_\mu$
and $\bar\nu_\mu$ without change of lepton number or flavor-lepton number.
Such pair processes certainly occur in the form of non-forward scattering
with a rate proportional to $\GF^2$, but can also occur on the refractive
level with a rate proportional to $\GF$. For most cases studied in the
literature, the conversion rate was actually found to be of order $\Delta
m^2/2E$ instead, i.e., driven by the frequency $\omega$. Another possible
driving frequency is the neutrino-neutrino interaction energy $\mu=\sqrt{2}
G_{\rm F}n_{\nu}$. The very definition of a ``dense'' neutrino gas is
precisely that $\mu\gg\omega$. However, this dominant scale cancels when the
neutrino and antineutrino angle distributions are too similar. On the other
hand, with sufficiently different angle distributions the conversion rate can
be driven by $\mu\gg\omega$, corresponding to much faster conversion.
Moreover, these fast conversions can exist even without any vacuum frequency
$\omega$ and thus in the absence of neutrino masses.

In general, therefore, self-induced flavor conversion---in the sense of
flavor reshuffling among modes---can occur without flavor mixing, provided
there exist fluctuations in flavor space to seed the unstable modes. One may
speculate that even quantum fluctuations of the mean-field quantities could
suffice as seeds. However, in practice ordinary neutrino oscillations driven
by their masses and mixing parameters exist, so disturbances in flavor space
to seed self-induced flavor conversion always exist even on the mean-field
level.

The main purpose of our paper is to present a few simple examples which
illustrate these general points and which are even more basic than those
presented by Sawyer. We begin in section~\ref{sec:beam} with the simplest
possible case, two colliding beams of neutrinos and antineutrinos, which
shows fast flavor conversion if we allow for inhomogeneities. In
section~\ref{sec:homogeneous} we also construct a homogeneous example,
consisting of four modes in the form of two beams intersecting with a
nonvanishing angle. We finally turn in section~\ref{sec:two-bulb} to the
example of a spherical source which emits neutrinos and antineutrinos with
different zenith-angle distributions in analogy to the schematic supernova
model proposed by Sawyer~\cite{Sawyer:2015dsa}. We conclude with a discussion
and outlook in section~\ref{sec:conclusion}.

\section{Colliding beams}
\label{sec:beam}

The current-current structure of the low-energy neutrino-neutrino interaction
implies that we need at least two different propagation directions to obtain
any effects at all. Therefore, the simplest possible example is an initially
homogeneous gas of neutrinos and antineutrinos, allowing only for two
opposing directions of motion, i.e., a system that is one-dimensional in
momentum space and that we can view as two colliding beams (figure
\ref{fig:beam-setup}). This type of simple model was recently used by several
groups to study the impact of spontaneously breaking various symmetries
\cite{Raffelt:2013isa, Hansen:2014paa, Chakraborty:2015tfa, Duan:2014gfa,
Abbar:2015mca, Mangano:2014zda, Mirizzi:2015fva}.

\subsection{Linearized equations of motion}

On the refractive level, the interacting neutrino system is best represented
in terms of the mean-field $\varrho_i$ for every momentum mode $i$. The
diagonal components of this matrix in flavor space are phase-space densities
of the different flavor states, whereas the off-diagonal elements represent
flavor correlations. Antineutrinos are represented by negative energies and
we use the ``flavor isospin convention,'' where the $\varrho$ matrices for
antineutrinos correspond to negative phase-space densities. The advantage is
that we do not need to distinguish explicitly between neutrino and
antineutrino modes. The $\varrho$ matrices of $N$ modes evolve
according to \cite{Sigl:1992fn}
\begin{equation}\label{eq:EOM1}
\I\(\partial_t+{\bf v}_i\cdot{\bm \nabla}\)\,\varrho_i
=\[{\sf H}_i,\varrho_i\]\,,
\quad\hbox{where}\quad
{\sf H}_i=\frac{{\sf M}^2}{2E_i}
+\mu\sum_{j=1}^N\(1-{\bf v}_i\cdot{\bf v}_j\)\varrho_j\,,
\end{equation}
where ${\sf M}^2$ is the matrix of neutrino mass-squares. We assume neutrinos
to be ultra-relativistic so that the velocities ${\bf v}_i$ are unit vectors
giving the directions of the individual modes. The neutrino-neutrino
interaction energy is $\mu=\sqrt{2}\GF n_\nu$ with the effective neutrino
density $n_\nu=\half(n_{\nu_e}+ n_{\bar\nu_e}-n_{\nu_x}-n_{\bar\nu_x})$. We
always consider two-flavor oscillations between $\nu_e$ and another flavor
$\nu_x$ which is a suitable combination of $\nu_\mu$ and $\nu_\tau$. These
conventions follow reference~\cite{Chakraborty:2014lsa} and are chosen such
that a fixed $\mu$ corresponds to a fixed density of $\nu_e$ plus
$\bar\nu_e$, even when we modify, for example, their relative abundance. In
much of the previous literature, instead either the number of $\nu_e$ or of
$\bar\nu_e$ was held fixed, but we here prefer a more symmetric definition.

\begin{figure}
\centering
\includegraphics[width=0.40\textwidth]{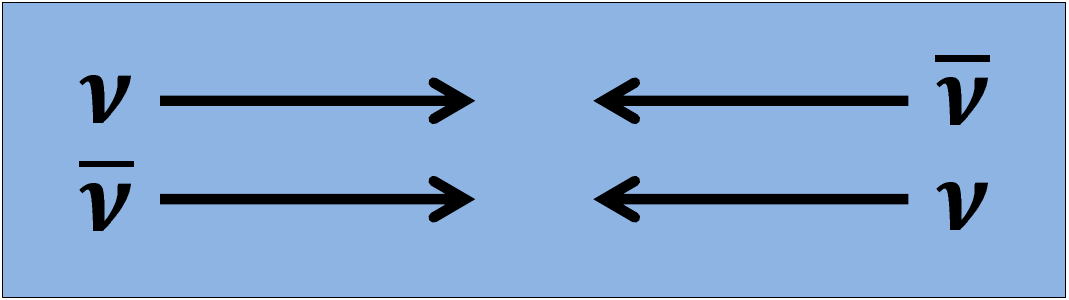}
\caption{Initially homogeneous ensemble of four neutrino modes (``colliding beams''
of neutrinos and antineutrinos). The system is taken to be infinite in all
directions. The normalized $\nu$ flux is $1+a$, the $\bar\nu$ flux
$1-a$ with the asymmetry parameter $a$ in the range
$-1\leq a\leq+1$. The left-right asymmetry is parametrized by $b$ such that
the upper beam in this figure has normalized strength $1+b$, the lower beam
$1-b$ with $-1\leq b\leq+1$. The relation of  parameters $r$, $\bar{r}$, $l$
and $\bar{l}$ to parameters $a$ and $b$ can  be found in equation
(\ref{Eq:param}).}\label{fig:beam-setup}
\end{figure}

In order to identify unstable modes, we consider a linearized version of
equation~(\ref{eq:EOM1}). We first note that ${\rm Tr}\,\varrho_i$ is
conserved by flavor conversion and, if the system was initially homogeneous,
it is not modified by the transport term in equation~(\ref{eq:EOM1}). It is
convenient to define traceless normalized $\varrho$ matrices in the form
\begin{equation}
\varrho_i-\frac{1}{2}\,{\rm Tr}\,\varrho_i
=\frac{g_i}{2}\begin{pmatrix}s_i&S_i\\ S_i^*&-s_i\end{pmatrix}\,,
\end{equation}
where $s_i$ is a real and $S_i$ a complex number with $s_i^2+|S_i|^2=1$.
Moreover, if neutrinos are initially prepared in $\nu_e$ or $\bar\nu_e$
eigenstates (our usual example), then initially $s_i=+1$. The ``spectrum''
$g_i$ gives the actual density of neutrinos in mode $i$ and is positive for
an initial $\nu_e$ and negative for an initial $\bar\nu_e$, corresponding to
our flavor-isospin convention. Our definition of the effective neutrino
density $n_\nu$ corresponds to the normalization $\sum_{i=1}^N|g_i|=2$. To
linear order, $s_i=1$ remains constant, whereas the off-diagonal elements
evolve according to
\begin{equation}\label{eq:EOM2}
\I\(\partial_t+{\bf v}_i\cdot{\bm \nabla}\)\,S_i=
\left[\omega_i+\mu\sum_{j=1}^N \(1-{\bf v}_i\cdot{\bf v}_j\)g_j\right]S_i
-\mu\sum_{j=1}^N\(1-{\bf v}_i\cdot{\bf v}_j\)g_j S_j\,.
\end{equation}
We have assumed a very small vacuum mixing angle and use $\omega_i=\Delta
m^2/2E_i$ with $\Delta m^2$ positive and the convention that $\omega_i$ is
positive for neutrinos and negative for antineutrinos.

As a next step, we transform this linear equation of the space-time
coordinates $(t,{\bf r})$ into Fourier space $(\Omega,{\bf k})$ and we write
$S_i(t,{\bf r})=Q_i(\Omega,{\bf k})\,e^{-\I(\Omega t-{\bf k}\cdot{\bf r})}$.
The linearized equations of motion in Fourier space are
\begin{equation}\label{eq:EOM3}
\Omega\,Q_i=
\left[\omega_i+{\bf v}_i\cdot{\bf k} +\mu\sum_{j=1}^N
\(1-{\bf v}_i\cdot{\bf v}_j\)g_j\right]Q_i
-\mu\sum_{j=1}^N\(1-{\bf v}_i\cdot{\bf v}_j\)g_j Q_j\,.
\end{equation}
We are looking for exponentially growing solutions, i.e., eigenvalues
$\Omega$ with a nonvanishing imaginary part.

We finally turn to a system which is one-dimensional in momentum space (a
beam) so that ${\bf v}_i\to v_i$ and ${\bf k}\to k$. The current-current
factors $(1-{\bf v}_i\cdot{\bf v}_j)$ are 0 for parallel-moving modes or 2
for opposite moving ones. We consider four modes as in
figure~\ref{fig:beam-setup} and use the vacuum oscillation frequency
$+\omega$ for neutrinos and $-\omega$ for antineutrinos. We denote the
amplitudes $Q_i$ for the different modes with $R$ for right-moving neutrinos,
$\bar R$ for right-moving antineutrinos, and $L$ and $\bar L$ for the left
movers. Likewise, we denote the mode occupations $g_i$ with $r$, $\bar r$,
$l$ and $\bar l$, respectively. Equation~(\ref{eq:EOM3}) becomes
\begin{equation}\label{eq:EOM4}
\Omega\,\begin{pmatrix}R\\ \bar L\\ L\\ \bar R\end{pmatrix}=\left[
\begin{pmatrix}\omega+k&0&0&0\\0&-\omega-k&0&0\\0&0&\omega-k&0\\0&0&0&-\omega+k\end{pmatrix}
+2\mu
\begin{pmatrix}
l+\bar l&-\bar l&-l&0\\
-r&r+\bar r&0&-\bar r\\
-r&0&r+\bar r&-\bar r\\
0&-\bar l&-l&l+\bar l\\
\end{pmatrix}
\right]\begin{pmatrix}R\\ \bar L\\ L\\ \bar R\end{pmatrix}\,.
\end{equation}
This is the most general one-dimensional four-mode case and the starting
point for the discussion in the rest of this section.

\subsection{Two modes}

The possible existence of unstable modes for $\omega=0$ is most easily
understood in a yet simpler case consisting only of right-moving neutrinos
and left-moving antineutrinos, i.e., $r=1+a$, $\bar l=-(1-a)$, and ${\bar
r}=l=0$. The parameter $-1\leq a\leq1$ codifies the neutrino-antineutrino
asymmetry of our system. Equation~(\ref{eq:EOM4}), reduced to the two
occupied modes, becomes
\begin{equation}\label{eq:EOM5}
\Omega\,\begin{pmatrix}R\\ \bar L\end{pmatrix}
=\left[\begin{pmatrix}\omega+k&0\\0&-\omega-k\end{pmatrix}+2\mu
\begin{pmatrix}-1+a,&1-a\\-1-a,&1+a\end{pmatrix}
\right]\begin{pmatrix}R\\ \bar L\end{pmatrix}\,.
\end{equation}
Without further calculation we can see that the role of the vacuum
oscillation frequency $\omega$ is here played by $\omega+k$. If we
consider vanishing neutrino masses ($\omega=\Delta m^2/2E=0$) and a
spatial Fourier mode $k>0$, the role usually played by $\omega$ will
be taken over by $k$. The reason for this behavior is that we have
constructed our system such that neutrinos (vacuum frequency
$+\omega$) move right so that the spatial Fourier term ${\bf
  v}\cdot{\bf k}$ enters as $+k$, and the other way round for the beam
of left-moving antineutrinos.

For completeness we provide the explicit eigenvalues for this two-mode case.
Using the notation $\tilde\omega=\omega+k$ we find
\begin{equation}
\Omega=2a\mu\pm\sqrt{(2a\mu)^2+\tilde\omega(\tilde\omega-4\mu)}\,.
\end{equation}
For $\tilde\omega=0$ the eigenvalues are purely real. The eigenfrequencies
have an imaginary part for
\begin{equation}
1-\sqrt{1-a^2}<\frac{\tilde\omega}{2\mu}<1+\sqrt{1-a^2}\,.
\end{equation}
Notice that in this system, the solutions for positive or negative $k$ are
different. Because $\mu$ is defined to be positive, we have unstable
solutions only for $\tilde\omega>0$, i.e., for $k>-\omega$. The
imaginary part has its maximum for $\mu=\tilde\omega/(2a^2)$ and is
\begin{equation}
{\rm Im}\,\Omega\vert_{\rm max}
=\tilde\omega\,\sqrt{\frac{1}{a^2}-1}\,.
\end{equation}
Therefore, in the homogeneous case ($k=0$), the growth rate is indeed
proportional to the vacuum oscillation frequency times a numerical factor
which depends on the neutrino-antineutrino asymmetry. Conversely, for
vanishing $\omega$, the maximum growth rate is proportional to the spatial
Fourier wave number $k$.

We can turn this discussion around and ask which Fourier modes $k$ are
unstable for a fixed $\mu$ value. The maximum growth rate occurs for
$\tilde\omega=2\mu$. The fastest-growing $k$ mode grows with the rate
\begin{equation}
{\rm Im}\,\Omega\vert_{\rm max}
=2\mu\,\sqrt{1-a^2}\,.
\end{equation}
This rate is indeed ``fast'' in the sense that it is proportional to
$\sqrt{2}\GF n_\nu$. Of course, we assume that $-1<a<1$ is not
fine-tuned to be very close to $\pm1$ which would correspond to having
only neutrinos or only antineutrinos in the system.

\subsection{Four modes}

The results of the previous section came about because the system was
constructed with maximal left-right asymmetry: neutrinos were only
right-moving, antineutrinos left-moving. On the other hand, previous one-beam
examples \cite{Raffelt:2013isa, Hansen:2014paa, Chakraborty:2015tfa, Duan:2014gfa,
Abbar:2015mca, Mangano:2014zda, Mirizzi:2015fva} had been constructed to be
left-right symmetric, although spontaneous symmetry breaking of
the solution was allowed. In all previous cases, the colliding beams were
stable against self-induced flavor conversion in the $\omega=0$ limit even
for nonvanishing $k$. Therefore, the system must be prepared with some amount
of left-right asymmetry to be unstable for vanishing $\omega$.

To study this condition, we now turn to a four-mode system with left- and
right-moving neutrinos and antineutrinos. As before, we use the parameter $a$
to denote the neutrino-antineutrino asymmetry, and in addition the parameter
$-1\leq b \leq 1$ to denote the left-right asymmetry. Specifically, we use the beam
occupations
\begin{subequations}\label{Eq:param}
\begin{eqnarray}
r     &=& +\half(1+a)(1+b)\,,\\
\bar l&=& -\half(1-a)(1+b)\,,\\
l     &=& +\half(1+a)(1-b)\,,\\
\bar r&=& -\half(1-a)(1-b)\,.
\end{eqnarray}
\end{subequations}
The two-mode example of the previous section corresponds to $b=1$. The
neutrino-neutrino interaction matrix in equation~(\ref{eq:EOM4})
becomes
\begin{equation}\mu
\begin{pmatrix}
2\,(a-b)&(1-a)(1+b)&~-(1+a)(1-b)&0\\
-(1+a)(1+b)&2\,(a+b)&~0&(1-a)(1-b)\\
-(1+a)(1+b)&0&~2\,(a+b)&(1-a)(1-b)\\
0&(1-a)(1+b)&~-(1+a)(1-b)&2\,(a-b)\\
\end{pmatrix}\,.
\end{equation}
The eigenvalue equation is now quartic and the explicit solutions provide
little direct insight.

However, in several special cases there are simple solutions. In the previous
literature, one always used a system which was set up in a left-right
symmetric configuration, meaning $b=0$. Considering the homogeneous case
($k=0$), one finds the eigenvalues
\begin{equation}
\Omega=a\mu\pm\sqrt{(a\mu)^2+\omega(\omega-2\mu)}
\quad\hbox{and}\quad
\Omega=3a\mu\pm\sqrt{(a\mu)^2+\omega(\omega+2\mu)}\,.
\end{equation}
The first solution corresponds to the usual ``flavor pendulum'' for inverted
neutrino mass ordering, the second to the left-right symmetry breaking
solution for normal ordering as discussed previously~\cite{Raffelt:2013isa}.
Notice that changing the mass ordering corresponds to ${\bf B}\to-{\bf B}$ in
equation~(\ref{eq:EOM1}) and thus to $\omega\to-\omega$ in these expressions.

The corresponding inhomogeneous case ($|k|>0$) was studied in
references~\cite{Duan:2014gfa, Chakraborty:2015tfa}. The system is stable for
$\omega=0$ and the growth rate is proportional to $\omega$. Explicit results
can be derived in the $k\to\infty$ limit~\cite{Chakraborty:2015tfa}.

Turning now to the left-right asymmetric case, one can actually find explicit
solutions for $\omega=0$,
\begin{equation}
\Omega=2a\mu\pm(k-2b\mu)
\quad\hbox{and}\quad
\Omega=2a\mu\pm\sqrt{(2a\mu)^2+k(k-4b\mu)}\,.
\end{equation}
All solutions are real in the homogeneous case ($k=0$) and for any $k$ in the
left-right symmetric system ($b=0$). In the general case, the second
eigenvalue has an imaginary part if the expression under the square root is
negative, which occurs for
\begin{equation}
b-\sqrt{b^2-a^2}
<\frac{k}{2\mu}<
b+\sqrt{b^2-a^2}\,.
\end{equation}
An instability exists only for $a^2<b^2$, i.e., the left-right asymmetry must
exceed the neutrino-antineutrino asymmetry. For fixed $\mu$, the Fourier mode
with the largest growth rate is $k=2b\mu$, growing with a rate
\begin{equation}
{\rm Im}\,\Omega\vert_{\rm max}
=2\mu\,\sqrt{b^2-a^2}\,.
\end{equation}
Again, this growth is of order a typical neutrino-neutrino interaction energy
and thus ``fast''.

\section{Intersecting beams}
\label{sec:homogeneous}

\begin{figure}[b]
\centering
\includegraphics[width=0.4\textwidth]{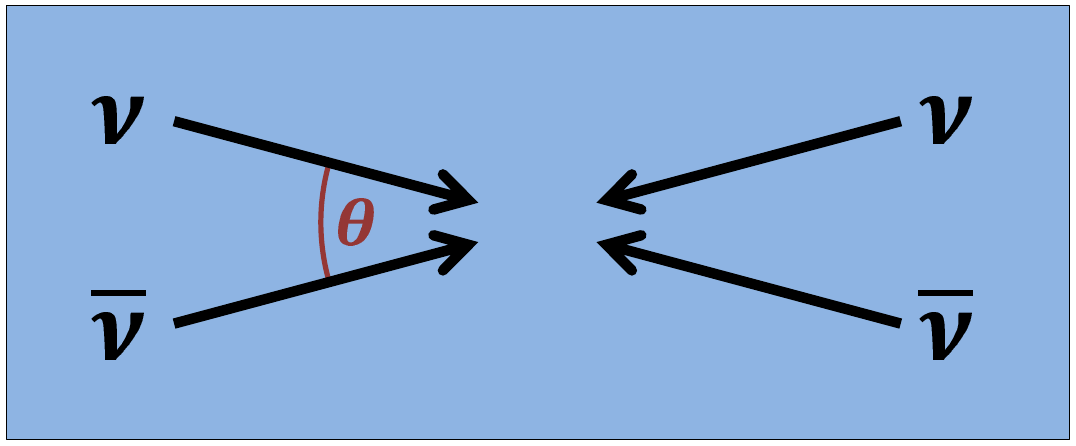}
\caption{Homogeneous ensemble of four neutrino modes (two beams with relative
angle $\theta$.) The system is taken to be infinite in all
directions.}\label{fig:twobeam}
\end{figure}

The crucial ingredient to obtain fast flavor conversion appears to be a
sufficient difference between the direction distribution of neutrinos and
antineutrinos. However, if the momentum distribution is one dimensional we
need spatial inhomogeneities. As a next step we construct the
simplest homogeneous system (${\bf k}=0$) that shows fast flavor conversion.
We consider four modes with directions which intersect at a relative angle
$\theta$ as shown in figure~\ref{fig:twobeam}. We continue to denote the
modes as ``left- and right-moving'' in an obvious sense. The mode occupations
are taken to be symmetric between left and right, but we include a
neutrino-antineutrino asymmetry $a$, i.e., the mode occupations are taken to
be
\begin{subequations}
\begin{eqnarray}
r=l          &=& +\half(1+a)\,,\\
\bar r=\bar l&=& -\half(1-a)\,.
\end{eqnarray}
\end{subequations}
As before, we use the normalization $|r|+|l|+|\bar r|+|\bar l|=2$. The
current-current factors $(1-{\bf v}_i\cdot {\bf v}_j)$ are equal to 2 for
opposite moving modes, and $1\pm\cos\theta$ for the other pairs in obvious
ways.

The symmetries of this setup suggest to combine the neutrino and antineutrino
amplitudes in a symmetric and antisymmetric form, $A_{\pm}=\half(L\pm R)$ and
$\bar{A}_\pm=\half(\bar L\pm \bar R)$. Indeed, the linearized equations of
motion decouple and we find with $c=\cos\theta$
\begin{eqnarray}
\kern-1em\Omega\,\begin{pmatrix}A_+\\ \bar A_+\end{pmatrix}&=&
\left[\begin{pmatrix}\omega&0\\ 0&-\omega\end{pmatrix}
+\frac{\mu\,(3-c)}{2}
\begin{pmatrix}
-1+a,&1-a,\\ -1-a&1+a
\end{pmatrix}
\right]\begin{pmatrix}A_+\\ \bar A_+\end{pmatrix},
\\[1.5ex]
\kern-1em\Omega\,\begin{pmatrix}A_-\\ \bar A_-\end{pmatrix}&=&
\left[\begin{pmatrix}\omega&0\\ 0&-\omega\end{pmatrix}
+\frac{a\mu\,(5+c)}{2}
+\frac{\mu}{2}
\begin{pmatrix}
-(1-3c)&-(1+c)(1-a)\\ (1+c)(1+a)&1-3c
\end{pmatrix}
\right]\begin{pmatrix}A_-\\ \bar A_-\end{pmatrix}.
\end{eqnarray}
The first equation again corresponds to the usual flavor pendulum. Indeed,
for $c=-1$ we return to the situation of a completely left-right asymmetric
system of all neutrinos moving right and all antineutrinos moving left and we
reproduce equation~(\ref{eq:EOM5}). As observed earlier, we then need a
nontrivial spatial variation to obtain fast flavor conversion.

\begin{figure}[b]
\centering
\includegraphics[width=0.7\textwidth]{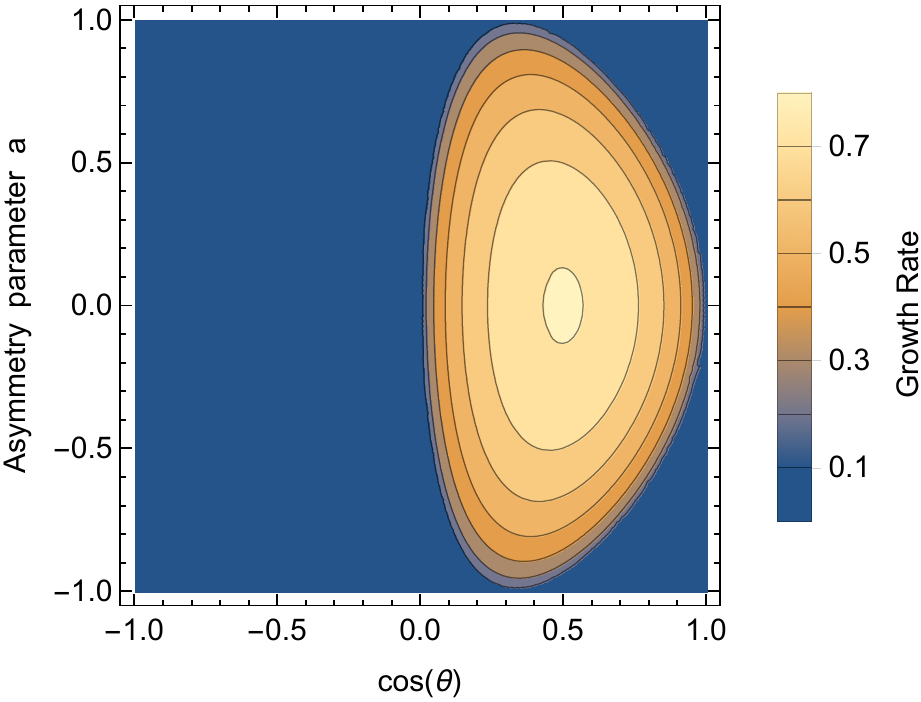}
\caption{Growth rate in units of $\mu$ for neutrino-antineutrino beams
  intersecting at an angle $\theta$ and a neutrino-antineutrino
  asymmetry $-1<a<1$. The mode occupations are taken to be symmetric
  between left and right. The analytic expression is given in
  equation~(\ref{eq:twobeam-eigenvalue}). A fast growth rate occurs
  for $\cos\theta>0$ and it is symmetric between positive and negative
  $a$.}
\label{fig:twobeam-growthrate}
\end{figure}

The second case with left-right symmetry breaking, on the other hand,
provides nontrivial eigenvalues even for $\omega=0$ which are found to be
\begin{equation}\label{eq:twobeam-eigenvalue}
\Omega=\frac{a\mu(5+c)}{2}\pm\frac{\mu}{2}\sqrt{(1+c)^2a^2-8c(1-c)}\,.
\end{equation}
In figure~\ref{fig:twobeam-growthrate} we show the imaginary part as a
contour plot in the $\cos\theta$--$a$ plane. A fast growth rate occurs only
for $\cos\theta>0$ and it is symmetric between positive and negative $a$. The
absolute maximum obtains for $a=0$ and $\cos\theta=\half$ and is found to be
${\rm Im}\Omega|_{\rm max}=\mu/\sqrt{2}$.

We have performed a numerical solution of the full nonlinear equations
for typical parameters $\cos\theta$ and $a$ in the unstable regime.
We find the usual behavior of an inverted oscillator. Given a small
perturbation, there is a long phase of exponential growth of the
transverse component, followed by a deep dip of the flavor content of
type $\nu_e\bar\nu_e\to\nu_x\bar\nu_x$ and back to $\nu_e\bar\nu_e$
and so on, similar to the usual flavor pendulum.

\clearpage

\section{Two-bulb supernova model}
\label{sec:two-bulb}

\subsection{Setting up the model}

\begin{figure}[b]
\centering
\includegraphics[width=0.45\textwidth]{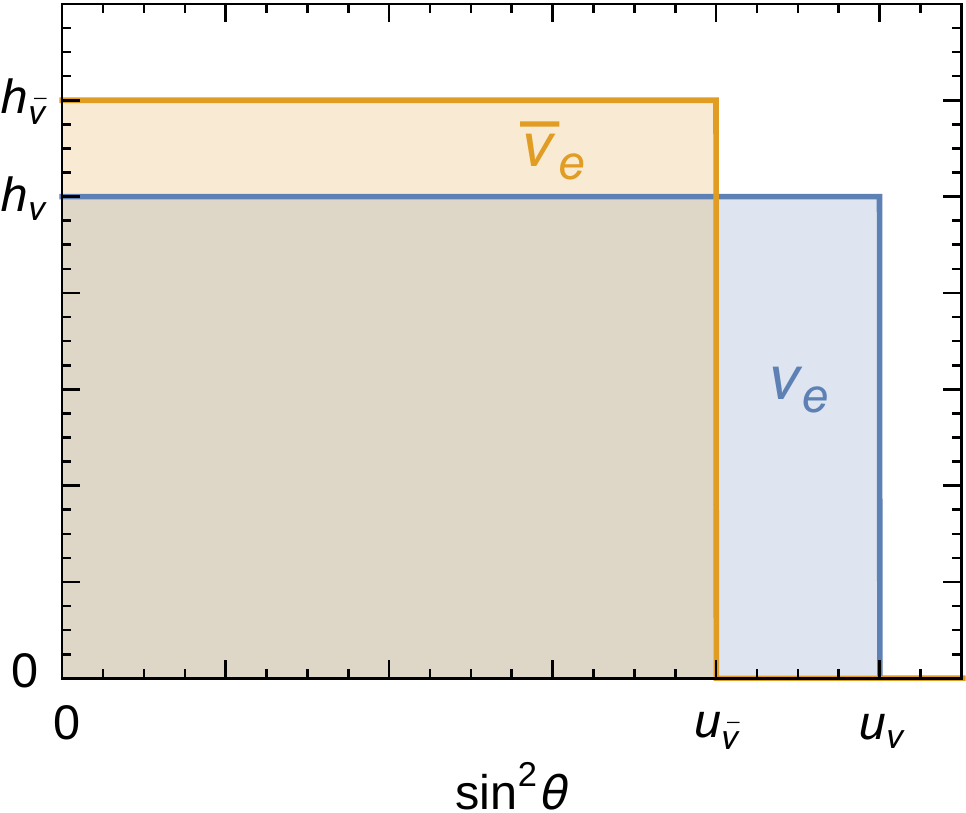}
\caption{Zenith-angle distribution for neutrinos (blue) and
antineutrinos (orange) implied by the two-bulb
supernova emission model.}
\label{fig:twobulb-spectrum}
\end{figure}

As a final example we consider the model proposed by
Sawyer~\cite{Sawyer:2015dsa} which is motivated by typical supernova emission
characteristics. Neutrinos are taken to emerge from a spherical surface, the
``neutrino bulb,'' with a blackbody-like angular characteristic, i.e.,
isotropically into the outer half space \cite{Duan:2006an}. In this case, a
distant observer measures a zenith-angle distribution which is uniform in the
variable $\sin^2\theta$ up to a maximum which is determined by the angular
size of the neutrino bulb at the observation point. The species $\nu_e$ and
$\bar\nu_e$ decouple in different regions. Therefore, as a simple
approximation one can assume that they are emitted from neutrino surfaces of
different radii, which we call a two-bulb emission model. This setup leads to
$\nu_e$ and $\bar\nu_e$ zenith-angle distributions of the type illustrated in
figure~\ref{fig:twobulb-spectrum}. In a supernova, one usually expects the
$\nu_e$ flux to exceed that of $\bar\nu_e$. However, the recently discovered
LESA phenomenon (lepton-emission self-sustained asymmetry) implies that the
relative fluxes show a strong hemispheric asymmetry \cite{Tamborra:2014aua}.
Moreover, in neutron-star mergers, very different distributions may occur
which also depend strongly on direction \cite{Dasgupta:2008cu,
Malkus:2014iqa}.

The main point of this supernova-motivated setup is the neutrino velocity
distribution in the transverse direction. One can formulate this problem in
terms of velocities in the transverse plane \cite{Chakraborty:2015tfa} and it
is then very similar to the colliding-beam examples of the previous sections,
with different velocity distributions for $\nu_e$ and $\bar\nu_e$. Therefore,
this case is conceptually quite similar to our previous ones.

We consider a stationary two-flavor neutrino flux and assume stationarity of
the solution, i.e., we study the flavor evolution as a function of radius. We
ignore small-scale effects in the transverse direction, i.e., the solution is
constrained to depend on radius alone. The neutrino radiation field at some
observation point beyond the emitting surface is described by the azimuth
angle $\varphi$ and the zenith-angle variable $u\propto\sin^2\theta$. The
range of occupied zenith angles is normalized to some chosen reference
radius, so the $u$-range does not depend on the test radius where we perform
the stability analysis. The emission spectrum $g(\omega,u)$ has the same
meaning as $g_i$ in our earlier sections, except that we use the continuous
labels $\omega$ and $u$. We assume axial symmetry of emission so that
$g(\omega,u)$ does not depend on $\varphi$.

\subsection{Eigenvalue equation}

We use the eigenvalue equation in the form developed in
reference~\cite{Raffelt:2013rqa} for the case of axially symmetric neutrino
emission. As input for the eigenvalue equation we need the integrals
\begin{equation}
I_n=\mu\int d\omega\,du\,\frac{u^n\,g(\omega,u)}{\omega+u\,\bar\lambda-\Omega}
\end{equation}
for $n=0$, 1 and 2. We have dropped a possible $\varphi$ dependence because
we assume axially symmetric emission. The parameter
$\bar\lambda=\lambda+\epsilon\mu$ describes the effective multi-angle matter
effect where $\lambda=\sqrt{2}\GF n_e$ and
\begin{equation}
\epsilon=\int d\omega\,du\,g(\omega,u)\,.
\end{equation}
In contrast to reference~\cite{Raffelt:2013rqa} we here normalize the
spectrum in the same way as in the previous sections, i.e., $\int
du\,d\omega\,{\rm sign}(\omega)\,g(\omega,u)=2$ which also influences the
meaning of $\mu$. The only physically relevant quantity is the product
$\mu\,g(u,\omega)$ and it is somewhat arbitrary how to normalize the two
quantities separately. The main point is to define a quantity $\mu$ which has
the meaning of a typical neutrino-neutrino interaction energy. The
eigenvalues $\Omega$ are found as solutions of one of the equations
\begin{equation}\label{eq:egnval}
(I_1-1)^2-I_0I_2=0
\qquad\hbox{and}\qquad
I_1+1=0\,.
\end{equation}
The first equation corresponds to those solutions which respect axial
symmetry, whereas the second corresponds to spontaneous axial symmetry
breaking.

We look for instabilities in the limit $\omega=0$.
In this case the contributions to $g(\omega,u)$ from emitted $\nu_x$ and
$\bar\nu_x$ drop out exactly if their emission characteristics are the same.
Notice, however, that the $\nu_x$ distribution enters indirectly through the
definition of the effective neutrino density $n_\nu$ and the definition of
$\mu$ and $\epsilon$. However, for simplicity we assume that no $\nu_x$ or
$\bar\nu_x$ are emitted. We denote the $\omega$-integrated zenith-angle
distributions as \smash{$h_{\nu_e}(u)=\int_0^\infty d\omega\,g(\omega,u)$}
for neutrinos (positive $\omega$) and
\smash{$h_{\bar\nu_e}(u)=-\int_{-\infty}^0 d\omega\,g(\omega,u)$} for
antineutrinos (negative $\omega$). In this notation we have
\begin{equation}
\int du\,\Bigl[h_{\nu_e}(u)+h_{\bar\nu_e}(u)\Bigr]=2
\qquad\hbox{and}\qquad
\int du\,\Bigl[h_{\nu_e}(u)-h_{\bar\nu_e}(u)\Bigr]=\epsilon\,.
\end{equation}
After performing the trivial $\omega$ integration, the above integrals are
\begin{equation}
I_n=\int du\,\frac{u^n}{u\,(\epsilon+m)-w}\,
\Bigl[h_{\nu_e}(u)-h_{\bar\nu_e}(u)\Bigr]\,,
\end{equation}
where $w=\Omega/\mu$ is the normalized eigenvalue and $m=\lambda/\mu$
represents the matter effect.

The two-bulb model of neutrino emission implies the top-hat $u$ distributions
shown in figure~\ref{fig:twobulb-spectrum}. We express the occupied
$u$-ranges in terms of a width parameter $-1<b<+1$ in the form
$u_{\nu_e}=1+b$ and $u_{\bar\nu_e}=1-b$. In the supernova context, the
$\nu_e$ interact more strongly, thus decouple at a larger distance, and hence
correspond to $b>0$. Moreover, we describe the normalized neutrino densities
as $n_{\nu_e}=1+a$ and $n_{\bar\nu_e}=1-a$ in terms of an ``asymmetry
parameter'' $-1<a<+1$. In the supernova context, deleptonization implies an
excess of $\nu_e$ over $\bar\nu_e$ so that $a>0$. In other words, the
traditional supernova-motivated situation corresponds to the first quadrant
$a,b>0$ in the parameter space of our model. In terms of these parameters,
the zenith-angle distributions are
\begin{equation}
h(u)=\frac{1\pm a}{1\pm b}\times\begin{cases}1&\hbox{for $0\leq u \leq 1\pm b$}\,,\\
0&\hbox{otherwise}\,,\end{cases}
\end{equation}
where the upper sign refers to $\nu_e$, the lower sign to $\bar\nu_e$. So
finally the integrals are
\begin{equation}
I_n=\frac{1+a}{1+b}\int_0^{1+b} du\,\frac{u^n}{u\,(2a+m)-w}
-\frac{1-a}{1-b}\int_0^{1-b} du\,\frac{u^n}{u\,(2a+m)-w}\,,
\end{equation}
where we have used $\epsilon=2a$. These integrals can be performed
analytically without problem, but the eigenvalues can be found only
numerically.

One special case is $b=0$ where the zenith-angle distributions for $\nu_e$
and $\bar\nu_e$ are the same, yet their number density is different
($a\not=0$). In this case the integrals are
\begin{equation}
I_n=2a\int_0^{1} du\,\frac{u^n}{u\,(2a+m)-w}\,.
\end{equation}
Numerically it appears that in this case the eigenvalues are always real,
i.e., fast flavor conversion indeed requires the top-hat distributions to
have different widths. However, we have not tried to prove this point
mathematically.

\subsection{Solution without matter effect}

\begin{figure}[b]
\centering
\includegraphics[width=0.7\textwidth]{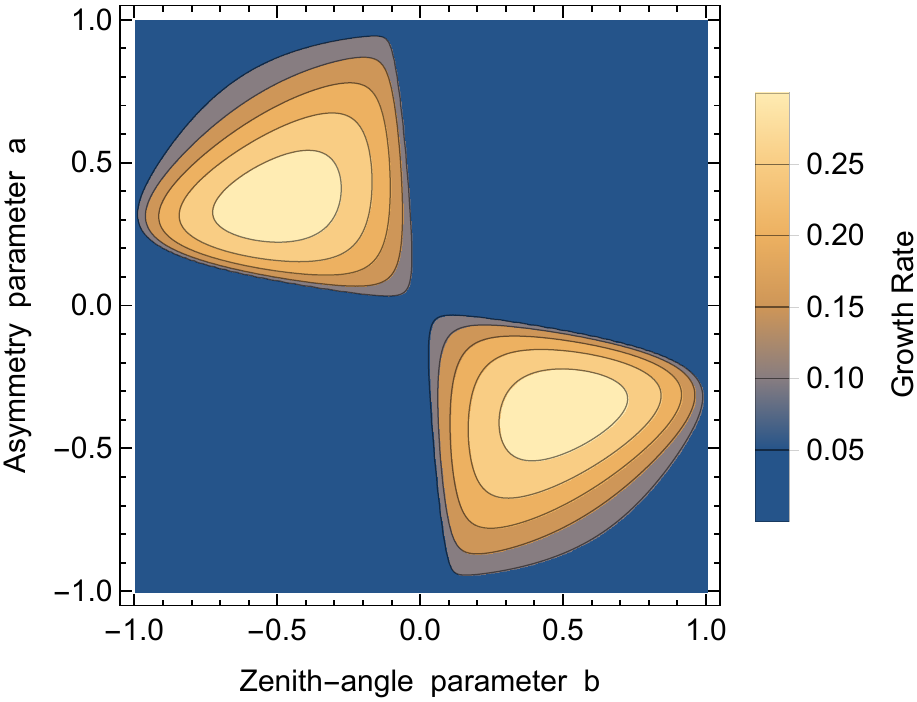}
\caption{Growth rate in units of $\mu$ for the axial-symmetry breaking
solution of the two-bulb supernova model without matter.
The normalized $\nu_e$ density is $1+a$, for $\bar\nu_e$ it is $1-a$. The
$\nu_e$ zenith-angle distribution is nonzero on the range $0<u<1+b$ and on
$0<u<1-b$ for $\bar\nu_e$. The results show no instability in the SN motivated parameters
($a>0$ and $b>0$), i.e., the first quadrant.
}\label{fig:twobulb-contour}
\end{figure}

As a first nontrivial case we ignore matter ($m=0$) and find that the first
case in equation~(\ref{eq:egnval}), the axially symmetric solution, does not
show any instabilities in a numerical scan over the space $-1<a<1$ and
$-1<b<1$. The second equation (axial symmetry breaking) provides solutions
and hence allows fast flavor conversion. We show the imaginary part of $w$,
i.e., the growth rate in units of $\mu$ as a contour plot in
figure~\ref{fig:twobulb-contour}. The first and third quadrants are stable,
i.e., when $a$ and $b$ have the same sign. These results suggest that fast
flavor conversion requires that the species $\nu_e$ or $\bar\nu_e$ with the
broader zenith-angle distribution must have a smaller flux. This particular
conclusion appears to be opposite of what Sawyer has found in his recent
study~\cite{Sawyer:2015dsa}.

Of course, in accretion disks arising from neutron-star mergers or black
hole-neutron star mergers, the flux is dominated by $\bar\nu_e$, not $\nu_e$,
so that $a<0$. Also LESA can be another interesting scenario spanning
parameters other than the traditional supernova-motivated case. Therefore, the main
point is the possible existence of fast flavor conversion if nontrivial
zenith-angle distributions are used.

\subsection{Including matter}

\begin{figure}
\centering
\includegraphics[width=0.7\textwidth]{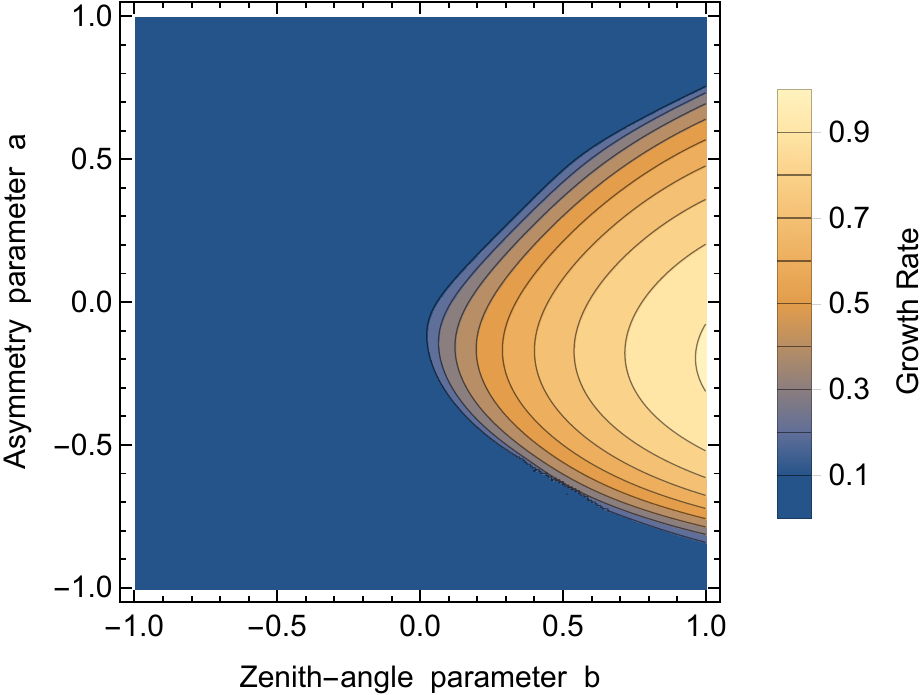}
\vskip20pt
\includegraphics[width=0.7\textwidth]{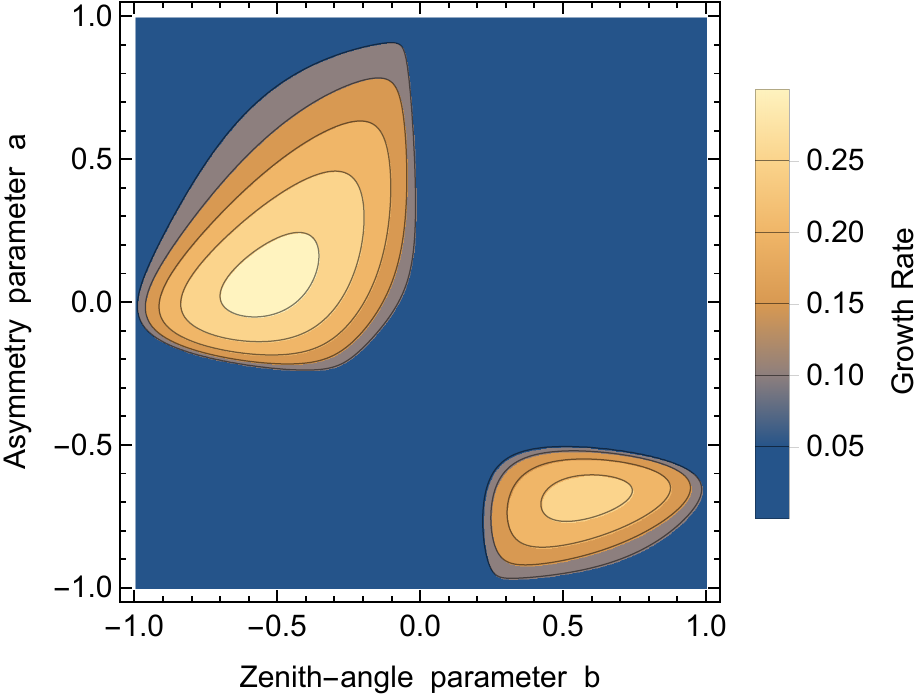}
\caption{Growth rate in units of $\mu$ in analogy to
figure~\ref{fig:twobulb-contour}, but now with matter $m=\lambda/\mu=1$. {\em
Top:\/} Axially symmetric solution. {\em Bottom:\/} Axial symmetry
spontaneously broken.}\label{fig:twobulb-matter}
\end{figure}

These above results change drastically in the presence of matter. A
substantial matter effect is expected when $\lambda$ is at least of order
$\mu$, so as a specific example we use $m=\lambda/\mu=1$ and show the growth
rates in figure~\ref{fig:twobulb-matter}. We find fast growth rates for both
the axially symmetric and the axial-symmetry breaking cases. While the latter
(bottom panel) is simply a modified version of the matter-free case, we now
find run-away solutions even in the axially symmetric case (upper panel). In
particular, there are unstable solutions for supernova-motivated parameters,
where the $\nu_e$ distribution is the broader one ($b>0$) and there are more
$\nu_e$ than $\bar\nu_e$ ($a>0$), i.e., the first quadrant of our parameter
space.

If we had used instead a background of antimatter ($m<0$), the unstable range
would lie in the other half where $b<0$. For $b=0$, when the two zenith-angle
distributions are the same, no fast instability seems to occur as remarked
earlier.

If the matter effect is very large ($\lambda\gg\mu$, corresponding to
$m\gg1$), the axially symmetric solution disappears, so it exists only for
some range of matter density. For example, during the supernova accretion
phase, this instability would be suppressed in analogy to the ``slow''
instabilities \cite{EstebanPretel:2008ni, Sarikas:2011am, Sarikas:2012vb,
Chakraborty:2011nf, Chakraborty:2011gd}. Of course, we have here only
considered the ${\bf k}=0$ case as well as stationarity of the solution.
Therefore, what all of this means in practice remains to be understood.

\subsection{Previous studies}

Flavor-dependent angle distributions were previously investigated by Mirizzi
and Serpico \cite{Mirizzi:2011tu, Mirizzi:2012wp}. These authors have not
reported fast flavor conversion in any of their cases. They have used
forward-peaked distributions of the form $(1-u)^{\beta/2}$ where $\beta=0$
provides a top-hat distribution on the interval $0< u<1$ and for $\beta>0$ a
distribution which is more concentrated for smaller $u$-values. However,
these authors assumed equal $\beta$ for both $\nu_e$ and $\bar\nu_e$ and a
different one for $\nu_x$ which was the same for $\bar\nu_x$, i.e., they only
studied angle differences between the $x$-flavor and the $e$-species. As we
remarked earlier, if $\nu_x$ and $\bar\nu_x$ have the same distribution, they
drop out of the equation in our limit of $\omega=0$. In other words, the
distributions used in references~\cite{Mirizzi:2011tu, Mirizzi:2012wp} indeed
do not spawn fast flavor conversion.

As a cross check we have also considered angle distributions of this form,
but taking different $\beta$ for $\nu_e$ and $\bar\nu_e$ as well as different
abundances. We find fast instabilities which qualitatively agree with our
earlier cases of top-hat distributions. Therefore, fast flavor conversion is
not an artifact of the top-hat distribution.

A stability analysis was also performed by Saviano et al.\
\cite{Saviano:2012yh}, using realistic energy and zenith-angle distributions
taken from a few specific numerical supernova simulations. The growth rates
reported in their figure~2 are always of order the vacuum oscillation
frequency and thus not fast. The used numerical angle distributions are shown
in the lower panels of their figure~1. The $\nu_e$ and $\bar\nu_e$ (dotted
and solid curves) look visually very similar except for the overall
normalization which represent the different fluxes. The matter effect was
taken into account, but not the possibility of axial symmetry breaking.
However, the fast conversion will remain absent when the ordinary matter effect is
significantly larger than the effect of background neutrinos.
Indeed, in the examples of reference
\cite{Saviano:2012yh} the matter effect $\lambda$ is almost an order of
magnitude larger than $\mu$. Also, probably in these specific models, the
angle distributions were not different enough to spawn fast flavor
conversion.
In a later study \cite{Chakraborty:2014nma}, these authors analyzed the same
models for the axial symmetry breaking case.
However, again the matter effects were large and the angle distributions
were similar. The lepton asymmetry was supernova inspired, i.e., the first quadrant
in figure~\ref{fig:twobulb-contour} and bottom panel of figure~\ref{fig:twobulb-matter}.
Unsurprisingly, fast flavor conversion did not show up in this case either.

In summary, while a number of previous studies have considered nontrivial
zenith-angle distributions, the chosen examples could not have found fast
flavor conversion.

\section{Conclusions}
\label{sec:conclusion}

We have studied a few simple examples of interacting neutrino systems which
show the phenomenon of ``fast flavor conversion,'' i.e., they have unstable
modes in flavor space which grow with rates of order the neutrino-neutrino
interaction energy $\mu=\sqrt{2} G_{\rm F}n_{\nu}$ instead of the much
smaller vacuum oscillation frequency $\omega=\Delta^2m/2E$. In these cases,
self-induced flavor conversion in the sense of flavor shuffling among modes
does not depend on $\Delta m^2$ or the vacuum mixing angle except for
providing disturbances as seeds for the run-away modes. In other words, the
main conceptual point is that self-induced flavor conversion does not depend
on flavor mixing. In the supernova context, neutrino flavor evolution on the
refractive level would have had to be considered even if flavor mixing among
neutrinos did not exist.

Notice that to lowest order, neutrino-neutrino interactions are of
neutral-current type and thus flavor blind. We ignore radiative corrections
which introduce a flavor dependence in neutrino-neutrino refraction
\cite{Mirizzi:2009td}. In this approximation, the overall flavor content of
the ensemble remains conserved by the action of neutrino-neutrino refraction,
i.e., self-induced flavor conversion corresponds to flavor reshuffling among
modes which however can lead to flavor decoherence if neighboring modes
become effectively uncorrelated.

The principle of fast flavor conversion was discovered ten years ago by Ray
Sawyer~\cite{Sawyer:2005jk} in a three-flavor setup of a small number of
modes. He speculated that supernova neutrinos might flavor-equilibrate over
very short distances, meters or even centimeters, in their decoupling region.
With hindsight it is difficult to understand why the conceptual and practical
points raised in this paper were completely lost on the community.

Fast flavor conversion by definition does not depend on the vacuum
oscillation frequencies and thus not on neutrino energy. The energy spectrum
plays no role, fast flavor conversion is driven by nontrivial angle
distributions. In several of our examples, the spontaneous breaking of
initial symmetries was also important. However, the crucial condition is that
the initial angle distribution must not be too symmetric or too simple,
although we cannot provide a general mathematical condition.

In the context of astrophysical applications in supernovae or neutron-star
mergers, the main question is if neutrinos emerging from the decoupling
region maintain spectral fluxes which strongly depend on species or if
self-induced flavor conversion and its interplay with matter effects and
vacuum oscillations leads to quick flavor decoherence. The effects of spatial
\cite{Duan:2014gfa, Abbar:2015mca, Mangano:2014zda, Mirizzi:2015hwa} and
temporal \cite{Abbar:2015fwa, Dasgupta:2015iia, Chakraborty:2016yeg} symmetry breaking as well as
the possibility of fast flavor conversion \cite{Sawyer:2005jk,
Sawyer:2015dsa} have been taken as evidence for quick decoherence. Still, the
breaking of spatial homogeneity may be suppressed by the multi-angle matter
effect \cite{Chakraborty:2015tfa}, and the breaking of stationarity depends on a
narrow resonance condition.

Actually, our stability studies as well as numerical solutions of the full
equations in the free-streaming limit may not be appropriate to capture the
realistic evolution at or near the neutrino decoupling region of a compact
object. In this region, the description of the neutrino mean field in terms
of a freely outward streaming neutrino flux is not appropriate: neutrinos
flow in all directions, but with different intensity. Even at larger
distances, the re-scattered neutrino flux plays an important role
\cite{Cherry:2012zw, Cherry:2013mv}.

Therefore, the toy examples studied here and in the recent literature leave
crucial questions open and do not yet provide clear-cut conclusions
concerning realistic flavor evolution in core-collapse or neutron-star merger
events. Neutrino flavor evolution in dense media remains a challenging
subject where different pieces of the jigsaw puzzle keep showing up, but do
not yet form a complete picture.

\section*{Acknowledgments}

We acknowledge partial support by the Deutsche Forschungsgemeinschaft through
Grant No.\ EXC 153 (Excellence Cluster ``Universe'') and by the
European Union through the Initial Training Network ``Invisibles,'' Grant
No.\ PITN-GA-2011-289442.

\begingroup\raggedright

\endgroup

\end{document}